\documentclass[11pt]{article}
\usepackage{eurosym}
\usepackage{amsmath}
\usepackage{amsfonts}
\usepackage{amssymb}
\usepackage{graphicx}
\usepackage{epstopdf}
\usepackage{cite}
\usepackage[usenames]{color}
\usepackage{multirow}

\providecommand{\U}[1]{\protect\rule{.1in}{.1in}}
\newcommand{\ba}{\begin{array}}
\newcommand{\ea}{\end{array}}

\newcommand{\Dsl}[1] { \setbox0=\hbox{$#1$}     
\dimen0=\wd0   \setbox1=\hbox{/} \dimen1=\wd1  \ifdim\dimen0>\dimen1        
 \rlap{\hbox to \dimen0{\hfil/\hfil}}  #1 \else \rlap{\hbox to \dimen1{\hfil$#1$\hfil}}  /  \fi  }
\newcommand{\bea}{\begin{eqnarray}}
\newcommand{\eea}{\end{eqnarray}}

\newcommand {\nb}{\bar n}

\begin{document}
\title{ {\Large  Relativistic corrections to $\psi(nS)\to\rho\pi$ exclusive decays and  their role in the understanding of  the $\rho\pi$-puzzle}  }
\author{ Nikolay Kivel \\
\textit{ Physik-Department, Technische Universit\"at M\"unchen,}\\
\textit{James-Franck-Str. 1, 85748 Garching, Germany } }
\maketitle

\begin{abstract}
We study relativistic corrections  to exclusive  $S$-wave charmonium decays into $\rho\pi$ and  $\gamma\pi$ final states.  The  contribution of relative order $v^2$ and the set of  associated higher order  corrections are calculated using NRQCD and collinear factorisation framework.  Numerical estimates show that the dominant effect is provided by the corrections of relative order $v^2$. The numerical values of these contributions are of the same order as the leading-order ones. 
These results suggest a scenario  where the sum of relativistic and radiative QCD corrections could explain the $\rho\pi$-puzzle. \end{abstract}

\newpage
\section{Introduction}
A description of  $S$-wave charmonium decays into $\rho\pi$ final state is already a long-standing problem in QCD phenomenology. The branching ratios for $J/\psi$ and excited state $\psi(3686)\equiv \psi'$  are measured  sufficiently accurately  and  their ratio is found to be very small \cite{Workman:2022ynf}
\bea
Q_{\rho\pi}=\frac{Br[\psi'\to \rho\pi]}{Br[J/\psi\to \rho\pi]}\approx 0.20\times 10^{-2}.
\label{Qrp}
\eea
This corresponds to a strong violation of the 13\%-rule,  which suggests that  $Q_{\rho\pi}\approx Q_{e^+e^-}\simeq 0.13$. 
The latter is  valid only if  the decay amplitudes of  $S$-wave charmonium  are dominated  by the leading-order contribution in the QCD factorisation framework (pQCD).  Therefore the disagreement between the data and  qualitative  theoretical  expectation indicates about large dynamical effects, which are not accounted by the  leading-order approximation of pQCD. 

The problem has attracted a lot of attention and many different qualitative ideas and phenomenological models have been proposed in order to understand the small value of $Q_{\rho\pi}$.  Almost all of  proposed explanations use different ideas about long distance QCD dynamics; a comprehensive overview of the topic  can be found in Refs.\cite{Brambilla:2004wf, Mo:2006cy}.

The dominant role   of  some  non-perturbative dynamics  is related to the fact  that the QCD helicity selection rule suppresses the valence contribution to  the decay amplitude. Therefore, it is necessary to take into account  for the one of  outgoing mesons  a  non-valence  component of the wave functions, which is suppressed  by additional power $\Lambda/m_c$.  However,  already long ago in Refs. \cite{Chernyak:1983ej, Zhitnitsky:1985bh} it was found that pQCD framework yields a reliable  leading-order estimate for  the $J/\psi$  branching ratio.   In Ref. \cite{Chernyak:1983ej} the non-valence contributions are described by the  three-particles twist-3  light-cone distribution amplitudes (LCDAs). These non-perturbative functions  are process independent and the first few moments of these functions can be estimated using QCD sum rules.  At present time corresponding  matrix elements were studied and  revised  for various mesons, see  updates in Refs. \cite{Ball:1998fj, Ball:2006wn, Ball:2007rt}.  Therefore it is reasonable to believe that pQCD description is a good starting point in order to develop  a  systematic description of the process within the effective field theory framework. 

Following  this way  one  faces with the problem in the description of  $\psi'\to  \rho\pi$, which  must be  strongly suppressed relative $J/\psi\to  \rho\pi$   in order to get  the small ratio (\ref{Qrp}).  There are various assumptions  about possible dynamical origins  for  this suppression.  Often they are related to the fact that the mass of excited state  $\psi'$ is close to the open charm threshold and this can lead to dynamical effects, which provide the crucial difference between $J/\psi$ and $\psi'$ decays.  The possible scenarios include:  destructive interference of the large non-valence and valence contributions  \cite{Chernyak:1983ej,Chernyak:1999cj};  suppression of the colour-singlet  $c\bar c$-wave function at the origin for $\psi'$ and the dominance of the colour-octet state \cite{Chen:1998ma}; cancellation between $c\bar c$ and $D\bar D$ components of $\psi'$ \cite{Suzuki:2000yq};   cancellation between $S$- and $D$-wave  components of $\psi'$ \cite{Rosner:2001nm} and others \cite{Brambilla:2004wf}.     

On the other hand, the potential of the effective field theory framework to study the problem was not been fully exploited yet.    Especially it is interesting to study  the  higher order corrections, which are  different  for $J/\psi$ and $\psi'$.  In this way the natural violation of the 13\% rule  can be related to  relativistic corrections in NRQCD \cite {Bodwin:1994jh}. 

In fact, already an order  $v^2$  nonrelativistic QCD matrix elements   
$\left\langle 0\right\vert\chi^{\dag} \boldsymbol{\sigma}\cdot\boldsymbol{\epsilon\ }
\left(  -{i}/{2}\overleftrightarrow{\boldsymbol{D}}\right)^{2}
\psi \left\vert \psi(n,\boldsymbol{ \epsilon})\right\rangle $  have very different values  for  $J/\psi$ and $\psi'$, that was  noticed already long time ago \cite{Braaten:2002fi}.
Recently, the relativistic corrections to exclusive  $\psi(nS) \to p\bar p$ decays have been studied  in Ref.\cite{Kivel:2022qjy}.  It is found that corrections of relative order $v^2$ are large and  comparable with the leading-order contribution.  This effect  is closely related to the structure of the integrand in the  collinear convolution integral describing the decay amplitude.  This observation  holds for both states  $J/\psi$  and $\psi'$ but  for excited state the absolute effect is larger  because   the  corresponding  matrix element is larger. The similar mechanism may also be relevant for other hadronic decay channels including  $\psi(nS) \to \rho\pi$ decays.   

Therefore, the main purpose of this paper is to calculate the relativistic corrections to $\psi(nS) \to \rho\pi$ and to  $\psi(nS) \to \gamma\pi$ decays  and to study their numerical effect. As a first step in this direction  we will  calculate  the correction of relative order $v^2$  combining NRQCD expansion with the leading-order collinear expansion.  We will use the NRQCD projection technique developed in Refs.\cite{Kuhn:1979bb, Bodwin:2002cfe,  Bodwin:2007fz, Bodwin:2007ga}, which is also effective  for calculations of  exclusive amplitudes.  This technique also allows one to resum  a part of higher order corrections, which are related to  the corrections to quark-antiquark wave function in the potential model \cite{Bodwin:2007ga}.  Such consideration  is also useful  providing  an estimate  of  possible effects from higher order contributions.

\section{Relativistic corrections to $\psi(nS)\to VP$  decays}

\subsection{ $\psi(nS)\to \rho\pi$ decay }

To describe the $\psi(nS,P)\rightarrow\rho(p)\pi(p^{\prime})$ decay amplitude
we use the charmonium rest frame and assume that outgoing  momenta are directed
along $z$-axis. The amplitude is defined as
\begin{equation}
\left\langle \rho(p)\pi(p')\right\vert iT\left\vert \psi(n, \boldsymbol{\epsilon})\right\rangle =i(2\pi
)^{4}\delta(p+p^{\prime}-P)i\epsilon_{\alpha\beta \mu\nu }%
\epsilon^{\alpha}e^{\ast\beta} \frac{p^{\prime\mu}p^\nu}{(pp^{\prime})}A_{\rho\pi},
\label{def:Apv}
\end{equation}
where $\epsilon$ and $e^{\ast}$ denotes polarisation vectors of  $\psi$
and $\rho$-meson, respectively.  
The amplitude $A_{\rho\pi}$ can be described as a superposition of a hard
kernel with  nonperturbative matrix elements describing the long distance coupling  with hadronic
states. In order to calculate the hard kernel, we  perform an NRQCD matching, 
which is combined with the  collinear light-cone expansion for the light quarks.  This technique allows one to perform the
matching at the amplitude level  and  to find the hard kernels for  corrections
associated with the specific set of higher order NRQCD matrix elements \cite{Bodwin:2007ga}
\begin{equation}
\left\langle {v}^{2n}\right\rangle =
\frac{
\left\langle 0\right\vert\chi^{\dag} \boldsymbol{\sigma}\cdot\boldsymbol{\epsilon\ }
\left(  -\frac{i}{2}\overleftrightarrow{\boldsymbol{D}}\right)^{2n}
\psi
\left\vert \psi(n,\boldsymbol{\epsilon})\right\rangle }
{m_{c}^{2n}\ \left\langle 0\right\vert \chi^{\dag}
\boldsymbol{\sigma}\cdot\boldsymbol{\epsilon}\psi\left\vert \psi(n,\boldsymbol{\epsilon})\right\rangle }\simeq\left\langle {v}^{2}\right\rangle^{n},
\end{equation}
where  the spatial part of the covariant derivative is defined as $\chi^{\dag}\overleftrightarrow{\boldsymbol{D}}\psi =\chi^{\dag}({\boldsymbol{D}}\psi)- ({\boldsymbol{D}}\chi^{\dag})\psi$ and  the last equality is valid up to corrections $\mathcal{O}(v^2)$ \cite{Bodwin:2006dn}.

\begin{figure}[ptb]%
\centering
\includegraphics[
width=3.5in
]%
{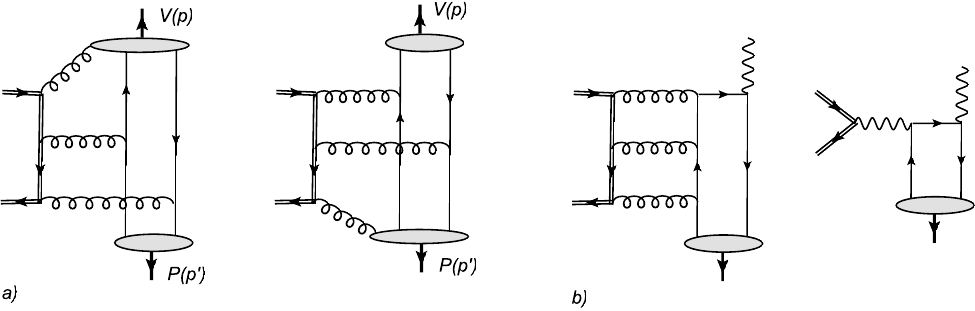}%
\caption{ a) Typical diagrams describing the subprocess $Q\bar{Q}\rightarrow
VP$, where $V=\rho, \gamma$. The blobs denote the light-cone matrix elements, see explanation in the
text.  b) An example of  diagrams, describing  the contribution with the  perturbative photon coupling. }%
\label{diagrams}%
\end{figure}

The diagrams, which describe the decay amplitude are schematically shown in
Fig.\ref{diagrams}.
The long distance hadronisation dynamics of outgoing mesons is described 
by the twist-2 and twist-3 light-cone distribution amplitudes (LCDAs).
Various properties and models for  required LCDAs can be
found in Refs.\cite{Ball:2006wn,Ball:2007rt}. The twist-2 light-cone matrix elements read \footnote{For simplicity, we do not  explicitly show the gauge links in the light-cone operators  assuming the appropriate light-cone gauge. }
\begin{equation}
\left\langle \pi^{+}(p^{\prime})\right\vert \bar{u}(z_{1+})\Dsl{\nb} \gamma
_{5}d(z_{2+})\left\vert 0\right\rangle =-if_{\pi}\left(  p^{\prime}\bar
{n}\right)  \ \int_{0}^{1}du\ e^{iu\left(  p^{\prime}\bar{n})(z_{1}n\right)
/2+i(1-u)\left(  p^{\prime}\bar{n})(z_{2}n\right)  /2}\ \phi_{2\pi}(u),
\label{tw2mepi}%
\end{equation}%
\begin{equation}
\left\langle \rho^{-}(p)\right\vert \bar{d}(z_{1-})\gamma_{\bot}^{\mu}%
\Dsl{n} u(z_{2-})\left\vert 0\right\rangle =if_{\rho}^{\bot}e_{\bot}^{\ast\mu
}(pn)\ \int_{0}^{1}dy\ e^{iy(pn)(z_{1}\bar{n})/2+i(1-y)\left(  pn)(z_{2}%
\bar{n}\right)  /2}\ \phi_{2\rho}^{\bot}(y), 
\label{tw2merho}%
\end{equation}
where  we use auxiliary light-cone vectors 
\begin{eqnarray}
n=(1,0,0,-1),\  \bar{n}=(1,0,0,1), \  g^{\mu\nu}_\bot=g^{\mu\nu}-\frac{1}{2}(n^\mu\bar{n}^\nu+n^\nu\bar{n}^\mu),
\\
p'=(p'\bar n)\frac{n}{2}+ \frac{m^2_\pi}{ (p'\bar n)}\frac{\bar n}{2} , \  p = (p n)\frac{\bar n}{2}+ \frac{m^2_\rho}{ (p n)}\frac{n}{2},\ (p'\bar n)\sim  (p n)\sim m_c.
\end{eqnarray}
and the short notation for the arguments of  quark
fields
\begin{equation}
q(z_{i+})\equiv q((z_{i}n)\bar{n}/2),\ q(z_{i-})\equiv\ q((z_{i}\bar
{n})n/2).
\end{equation}
The required twist-3 three-particles LCDAs  are defined as 
\begin{equation}
\left\langle \pi^{+}(p^{\prime})\right\vert \bar{u}(z_{1+})\Dsl{\nb}\gamma
_{\bot}^{\mu}\gamma_{5}gG_{\bar{n}\mu}(z_{3+})d(z_{2+})\left\vert
0\right\rangle =-2f_{3\pi}\left(  p^{\prime}\bar{n}\right)  ^{2}%
\text{FT}\left[  \phi_{3\pi}(u_{i})\right]  , \label{tw3mepi}%
\end{equation}
\begin{equation}
\left\langle \rho^{-}(p,e)\right\vert \bar{d}(z_{1-})\Dsl{n} gG_{\mu n}%
(z_{3-})u(z_{2-})\left\vert 0\right\rangle =-\ f_{\rho}m_{\rho}(pn)^{2}%
e_{\bot}^{\ast\mu}\ \text{FT}\left[  \phi_{3\rho}(y_{i})\right]  ,
\label{tw3merho}%
\end{equation}%
\begin{equation}
\left\langle \rho^{-}(p,e)\right\vert \bar{d}(z_{1-})\Dsl{n} \gamma_{5}g\tilde
{G}_{\mu n}(z_{3-})u(z_{2-})\left\vert 0\right\rangle =-if_{\rho}m_{\rho}%
\zeta_{3}(pn)^{2}e_{\bot}^{\ast\mu}\ \text{FT}\left[  \tilde{\phi
}_{3\rho}(y_{i})\right]  , \label{tw3merhot}%
\end{equation}
where for the gluon field strength tensor we use short notation $G_{\mu n}= G_{\mu\nu }n^\nu$. The dual gluon field strength tensor is defined as $\tilde{G}_{\mu \nu}=1/2\varepsilon_{\mu\nu\rho\sigma} {G}_{\rho \sigma}$ ( the Levi-Civita tensor is defined as $\varepsilon_{0123}=1$).   The symbol ``FT'' denotes the  Fourier transformation 
\begin{equation}
\text{FT}\left[  f(u_{i})\right]  =\int Du_{i}\ e^{iu_{1}\left(  p^{\prime
}\bar{n})(z_{1}n\right)  /2+iu_{2}\left(  p^{\prime}\bar{n})(z_{2}n\right)
/2+iu_{3}\left(  p^{\prime}\bar{n})(z_{3}n\right)  /2}f(u_{1}%
,u_{2},u_{3}), \label{def:FT}%
\end{equation}%
with
\begin{equation}
Du_{i}=du_{1}du_{2}du_{3}\delta(1-u_{1}-u_{2}-u_{3}).
\end{equation}
The FT$\left[  \phi_{3\rho}(y_{i})\right]  $ is defined analogously but
with $y_{i}(pn)(z_{i}\bar{n})$ in the Fourier transformation. The
normalisation constants $f_{\pi,\rho}$, $\zeta_{3},\ f_{3\pi}$ and models
for various LCDAs will be discussed below.

The expression for the amplitude can be written as
\begin{equation}
A_{\rho\pi}=\left\langle 0\right\vert \chi^{\dag}\boldsymbol{\sigma}%
\cdot\boldsymbol{\epsilon}\psi\left\vert \psi(n,\boldsymbol{\epsilon})\right\rangle
\frac{\sqrt{2M_{\psi}}}{2E}\ \frac{1}{4\pi}\int d\Omega\ \text{Tr}\left[
\Pi_{1}\hat{A}_{Q}\right]  , \label{Arpi}%
\end{equation}
where $\hat{A}_{Q}$ describes subprocess $Q\bar{Q}\rightarrow\rho\pi$ with the
quark-antiquark pair in the initial state.  The  heavy quark projector on the triplet spin state $\Pi_{1}$ reads \cite{Bodwin:2007ga}
\begin{equation}
\Pi_{1}=\frac{-1}{2\sqrt{2}E(E+m)}\left(  \frac{1}{2}\Dsl{P}  +m+\Dsl{q}
\right)  \frac{\Dsl{P}  +2E}{4E}\Dsl \epsilon \left(  \frac{1}{2}\Dsl{P}
-m-\Dsl{ q}  \right)  \otimes\frac{\mathbf{1}}{\sqrt{N_{c}}},
\end{equation}
and  is normalised as
\bea
\text{Tr}\left[  \Pi_{1}\Pi_{1}^{\dag}\right]  =4E^{2},
\eea
where $E$ is the heavy quark energy $p_{Q}=(E,\boldsymbol{q})$, $p_{\bar{Q}%
}=(E,-\boldsymbol{q})$ and $E=\sqrt{m_{c}^{2}+\boldsymbol{q}^{2}}$.  The
integration $d\Omega$ over the angles of the relative momentum $\boldsymbol{q}%
$ in Eq.(\ref{Arpi}) is used to get the state with $L=0$. Therefore the
relevant amplitude $\hat{A}_{Q}$ is the function of relative momentum square
$\boldsymbol{q}^{2}$ only, which  is
substituted $\boldsymbol{q}^{2}\rightarrow m_{c}^{2}\left\langle
{v}^{2}\right\rangle $ in the final expression (\ref{Arpi}), 
various technical details concerning NRQCD matching
can be found in Refs.\cite{Bodwin:2002cfe, Bodwin:2007ga}.

Calculation of  the diagrams as in Fig.\ref{diagrams} gives
\begin{equation}
A_{\rho^{-}\pi^{+}}=-\left\langle 0\right\vert \chi^{\dag}\boldsymbol{\sigma
}\cdot\boldsymbol{\epsilon}\psi\left\vert \psi(n, \boldsymbol{\epsilon})\right\rangle
\sqrt{2M_{\psi}}\left(  \pi\alpha_{s}\right)  ^{2}\frac{10}{27}\left(
1+\frac{m_{c}}{E}\right)  \frac{f_{\rho}^{\bot}~f_{3\pi}}{\left[
4E^{2}\right]  ^{2}}\left(  J_{\pi}+\frac{\ f_{\rho}m_{\rho}\zeta_{3}\ f_{\pi
}}{f_{\rho\bot}f_{3\pi}}\ J_{\rho}\right)  ,
\label{AVP}%
\end{equation}
where the dimensionless collinear convolution integrals $J_{\pi}$ and
$J_{\rho}$ describe contributions with twist-3  $\pi$- and $\rho$-LCDAs,
respectively. These integrals also depend on the  NRQCD parameter
$\left\langle {v}^{2}\right\rangle $. In the leading-order limit
$\left\langle {v}^{2}\right\rangle \rightarrow0$, $E\rightarrow
m_{c}^{2}$  Eq.(\ref{AVP}) reproduces the result from Ref.\cite{Chernyak:1983ej}
\begin{equation}
A_{\rho^{-}\pi^{+}}^{\text{lo}}=-\left\langle 0\right\vert \chi^{\dag
}\boldsymbol{\sigma}\cdot\boldsymbol{\epsilon}\psi\left\vert \psi(n,\boldsymbol{\epsilon})
\right\rangle \sqrt{2M_{\psi}}\left(  \pi\alpha_{s}\right)  ^{2}%
\frac{20}{27}\frac{f_{\rho}^{\bot}f_{3\pi}}{\left[  4m_{c}^{2}\right]  ^{2}%
}\left(  J_{\pi}^{\text{lo}}+\frac{\ f_{\rho}m_{\rho}\zeta_{3} f_{\pi}%
}{f_{\rho\bot}f_{3\pi}}\ J_{\rho}^{\text{lo}}\right)  ,\label{Alo}%
\end{equation}
where
\begin{align}
J_{\pi}^{\text{lo}} &  =\int Du_{i}\ \frac{\phi_{3\pi}(u_{i})}{u_{1}u_{2}%
u_{3}}\int_{0}^{1}dy\ \frac{\phi_{2\rho}^{\bot}(y)}{1-y}\frac{2u_{1}}{\left(
y\bar{u}_{2}+u_{2}\bar{y}\right)  \left(  yu_{1}+\bar{y}\bar{u}_{1}\right)
},\\
J_{\rho}^{\text{lo}} &  =\int_{0}^{1}du\ \frac{\phi_{2\pi}(u)}{1-u}\int
Dy_{i}\ \frac{\left(  \phi_{3\rho}+\tilde{\phi}_{3\rho}\right)  (y_{i})}%
{y_{1}y_{2}y_{3}}\frac{1}{y_{2}\bar{u}+u\bar{y}_{2}},
\end{align}
with $\bar{x}=1-x$. $\ $The analytical
expressions for the integrals $J_{\pi,\rho}$ in Eq.(\ref{AVP}) are somewhat
lengthy and presented in Appendix.

In order to estimate these integrals we use the following models of  LCDAs%
\begin{equation}
\phi_{2\rho}^{\bot}(y)=6y(1-y)\left(  1+a_{2\rho}\ C_{2}^{3/2}(2y-1)\right)
,\ \label{phi2rho}%
\end{equation}%
\begin{equation}
\phi_{2\pi}(u)=6u(1-u)\left(  1+a_{2}^{\pi}C_{2}^{3/2}(2u-1)\right)  ,
\label{phi2pi}%
\end{equation}%
\begin{equation}
\phi_{3\rho}(y_{i})=360y_{1}y_{2}y_{3}^{2}(y_{1}-y_{2})\omega_{3\rho},
\label{phi3rho}%
\end{equation}%
\begin{equation}
\tilde{\phi}_{3\rho}(y_{i})=360y_{1}y_{2}y_{3}^{2}\left(  1+\frac
{\tilde{\omega}_{3\rho}}{\zeta_{3}}\frac{1}{2}\left(  7y_{3}-3\right)
\right)  , \label{phi3rhot}%
\end{equation}%
\begin{equation}
\ \ \phi_{3\pi}(u_{i})=360u_{1}u_{2}u_{3}^{2}\left(  1+\omega_{3\pi}\frac
{1}{2}(7\alpha_{3}-3)\right)  . \label{phi3pi}%
\end{equation}
The different nonperturbative moments, which enter in the definitions
(\ref{tw2mepi})-(\ref{tw3merhot}) and (\ref{phi2rho})-(\ref{phi3pi}),
were estimated  in Refs.\cite{Chernyak:1983ej, Ball:2006wn, Ball:2007rt}.  Their values are
summarised in Table \ref{LCDAmom}. In the numerical estimates we fix for the factorisation
scale the value $\mu=2$ GeV and use $\alpha_{s}\simeq 0.30$.
\begin{table}[th]
\caption{ 
The values of the  moments, which parametrise the  hadronic LCDAs. 
All values are given at the scale $\mu=2$ GeV. For the pion moments, the values are taken from Ref.\cite{Ball:2006wn}, for the $\rho$-meson from Ref.\cite{Ball:2007rt}.
}
\label{LCDAmom}
\begin{center}
\begin{tabular}
[c]{|c|c|c|c|c|c|c|c|c|c|}\hline
$f_{\pi},$MeV & $f_{\rho}$, MeV & $f_{\rho}^{\bot}$, MeV & $a_{2\pi}$ &
$a_{2\rho}$ & $f_{3\pi},$ GeV$^{2}$ & $\zeta_{3\rho}$ & $\omega_{3\rho}$ &
$\tilde{\omega}_{3\rho}$ & $\omega_{3\pi}$\\\hline
$131$ & $216$ & $143$ & $0.19$ & $0.11$ & $0.31\times10^{-2}$ & $0.02$ &
$0.09$ & $-0.04$ & $-1.1$\\\hline
\end{tabular}
\end{center}
\end{table}

All the convolution integrals calculated with the models (\ref{phi2rho}%
)-(\ref{phi3pi}) are well defined, which confirms  that collinear
factorisation  is also valid beyond the leading-order approximation.

As a first step of the numerical analysis let us consider the the
leading-order estimate for the branching ratio of $J/\psi$. For that purpose
we use the  estimates for the NRQCD matrix element obtained in Ref.\cite{Bodwin:2007fz} 
\begin{equation}
\left\vert \left\langle 0\right\vert \chi^{\dag}\boldsymbol{\sigma}%
\cdot\boldsymbol{\epsilon}\psi\left\vert J/\psi\right\rangle
\right\vert ^{2}\simeq 0.440 \text{ GeV}^3.
\end{equation}
For the various masses in Eq.(\ref{Alo}) we use $M_{\psi}=3.1\ $GeV,
$m_{\rho}=775\ $MeV,  for the pole $c$-quark mass  $m_{c}=1.4\ $GeV and  for the total width $\Gamma_{J/\psi
}=93\ $KeV \cite{Workman:2022ynf}.  Then for the sum of all final states $\rho^{\pm
}\pi^{\mp}$ and $\rho^{0}\pi^{0}$ we obtain
\begin{equation}
\text{Br}[J/\psi\rightarrow\rho\pi]_{\text{lo}}\simeq1.0\%,
\end{equation}
which is somewhat smaller then the corresponding experimental value
$1.69(15)\%.$ This updated result confirm the conclusion of Ref.\cite{Chernyak:1983ej},
that the LO NRQCD approximation works sufficiently well  for the $J/\psi$
decay.\footnote{We assume that the difference about factor two is not a large discrepancy taking into account various uncertainties from scale setting, pole mass $m_c$, etc., which we do not consider now. } On the other hand this approximation can not describe branching ratio $\psi'\to\rho\pi$. 

Consider now the effect provided by the relativistic corrections in
Eq.(\ref{AVP}). The one part is provided by the resummation of relativistic
corrections in the factor $E=m_{c}^{2}\sqrt{1+\left\langle {v}%
^{2}\right\rangle }$ in Eq.(\ref{AVP}). This effect can be understood as
transition from the scale $4m_{c}^{2}$ to the scale $M_{\psi}^{2}\simeq
4m_{c}^{2}(1+\left\langle {v}^{2}\right\rangle )$. These
corrections reduce the ratio $Q_{\rho\pi}$  due to the factor \ $(1+\left\langle
{v}^{2}\right\rangle _{J/\psi})^{2}/(1+\left\langle {v}%
^{2}\right\rangle _{\psi^{\prime}})^{2}\sim M_{\psi}^{4}/M_{\psi^{\prime}}%
^{4}\sim0.50$. However, this can not explain the very small value
$Q_{\rho\pi}$ in Eq.(\ref{Qrp}).  

The second effect of the relativistic corrections is associated with the
modification of the hard kernels in the convolution integrals $J_{\rho,\pi}$.
Because these integrals depend on meson LCDAs, the resulting effect of the
relativistic corrections is also sensitive to hadronic nonperturbative
structure.  

For the numerical calculation we use for $J/\psi$ the estimate from Ref.\cite{Bodwin:2007fz}
\begin{equation}
\left\langle {v}^{2}\right\rangle _{J/\psi}\approx0.225,
\end{equation}
and for the excite state $\psi'$ we apply the following estimate%
\begin{equation}
\left\langle {v}^{2}\right\rangle _{\psi^{\prime}}=\frac
{M_{\psi^{\prime}}-M_{J/\psi}+E_{1}}{m_{c}}\approx0.64, \label{v2p}%
\end{equation}
where $E_{1}=\left\langle {v}^{2}\right\rangle _{J/\psi} m_c \simeq 315\ $MeV  is the binding energy for $J/\psi$.
The resulting value of $\left\langle {v}^{2}\right\rangle _{\psi^{\prime}}$ is much larger than  $\left\langle {v}^{2}\right\rangle_{J/\psi}$, which can have a significant numerical effect and, therefore, affect the value of $Q_{\rho\pi}$.

The given calculation of the relativistic corrections is complete at the
relative order $v^{2}$ only. The resummation of higher orders $\left\langle
{v}^{2}\right\rangle ^{n}$ with $n>2$ describes the part of the
relativistic corrections  associated with the quark-antiquark wave function
only \cite{Bodwin:2007ga}.  We use this approximation in order  to study
a possible effect from higher-order contributions. Therefore, for the
comparison, we present the values of the integrals in Eq.(\ref{AVP}) obtained
in  the leading-order approximation  $J^{\text{lo}}$ ($\left\langle
{v}^{2}\right\rangle \rightarrow0$),  in the next-to-leading
approximation $J^{\text{nlo}}$, which takes into account the next-to-leading
correction 
\begin{equation}
J^{\text{nlo}}=J^{\text{lo}}+\left\langle {v}^{2}\right\rangle J^{(1)},
\label{def:Jnlo}
\end{equation} 
and the integral $J$, which includes all powers $\left\langle {v}^{2}\right\rangle^n$
\begin{equation}
 J=J^{\text{lo}}+\sum\left\langle {v}^{2}\right\rangle ^{n}J^{(n)}.
\end{equation}

The total integral in Eq.(\ref{AVP}) is described by the sum of two
contributions with the different LCDAs
\begin{equation}
J_{\rho\pi}=J_{\pi}+\frac{\ f_{\rho}m_{\rho}\zeta_{3}\ f_{\pi}}{f_{\rho\bot
}f_{3\pi}}\ J_{\rho},\label{def:J}%
\end{equation}
where, schematically, $J_{\pi}=\phi_{3\pi}\ast T_{\pi}\ast\phi_{2\rho}^{\bot}$
and $J_{\rho}= \phi_{3\rho}\ast T_{\rho} \ast\phi_{2\pi}+\tilde{\phi}_{3\rho}\ast
\tilde{T}_{\rho} \ast\phi_{2\pi}$ (the asterisk denotes the
convolution integrals, $T_{\pi,\rho}$ are the hard kernels). Using parameters from Table~\ref{LCDAmom} one  finds
\begin{equation}
\frac{\ f_{\rho}m_{\rho}\zeta_{3}\ f_{\pi}}{f_{\rho\bot}f_{3\pi}}%
\approx0.99.\label{Rrho}%
\end{equation}
Therefore the normalisation couplings in the definitions (\ref{tw2mepi}%
)-(\ref{tw3merhot}) do not provide any numerical difference between the two
terms in Eq.(\ref{def:J}).  The results for  convolution integrals
(\ref{def:J}) are presented in Table~\ref{Jrp}
\begin{table}[th]
\caption{ Numerical result for the convolution integrals $J_{\rho\pi}$}
\label{Jrp}
\begin{center}
\begin{tabular} [c]{|c|c|c|c|}
\hline
& $J_{\rho\pi}^{\text{lo}}$ & $ J_{\rho\pi}^{\text{nlo}}/J_{\rho\pi
}^{\text{lo}}$ & $\ J_{\rho\pi}/J_{\rho\pi}^{\text{lo}}$\\\hline
$J/\psi$ & $630$ & $0.53$ & $0.45$\\\hline
$\psi^{\prime}$ & $630$ & $-0.46$ & $-0.65$\\\hline
\end{tabular}
\end{center}
\end{table}

The effect of the relativistic corrections is negative and the values of the LO
integrals are substantially reduced.  Notice that neglecting the higher-order corrections in $v^2$ in the square of the integral,  one gets in case of $J/\psi$  the  strong cancellation 
 \bea
|J_{\rho\pi}^{\text{nlo}}|^2 = (J_{\rho\pi}^{\text{lo}})^2
(2J_{\rho\pi}^{\text{nlo}}/J_{\rho\pi}^{\text{lo}} -1)+\mathcal{O}(v^4)\simeq 0.06 (J_{\rho\pi}^{\text{lo}})^2.
\eea
Therefore we assume that it is better to take the large NLO correction  exactly, i.e. do not expanding  the square of the integral in powers of $v^2$.  At the same time the numerical effect from other higher order  corrections is already much smaller. 

 For $\psi^{\prime}\to \rho\pi$ the numerical effect is  bigger  because  $\left\langle \boldsymbol{v}^{2}%
\right\rangle _{\psi^{\prime}}$ is larger. One can also see that the dominant part
of the correction is also provided by the contribution of relative order $v^2$, which is obtained exactly
in this calculation.  The numerical dominance of this correction can be explained by the numerical enhancement of the corresponding  convolution integrals  in the same way as for the baryon decays \cite{Kivel:2022qjy}.  

Let us assume that the  relativistic  correction of order $v^2$ provides the dominant numerical effect for $J/\psi$ and
 $\psi^{\prime}$ states. Then,  this allows one to suggest a possible explanation of
the small  $\psi^{\prime}\to \rho\pi$ width, which could explain the $\rho\pi$-puzzle.

 The NRQCD description of decay amplitudes also involves the $\mathcal{O}(\alpha_{s})$
NLO QCD radiative correction, which can also provide a substantial numerical
effect. Usually this contribution is considered to be of the same order as
relativistic corrections of relative order $v^{2}$.  In this case  the total convolution  integral   $J_{\rho\pi}$
to the next-to-leading accuracy is  given by, see Eq.(\ref{def:Jnlo})\footnote{ For simplicity, we show only the relative power of the QCD coupling $\alpha_s$ }
\begin{equation}
J^{\text{nlo}}_{\rho\pi}= J^{\text{lo}}+\left\langle {v}^{2}\right\rangle J^{(1)}+\alpha_s I^{(1)}
\end{equation}
where the integrals $J^{(1)}$ and $I^{(1)}_{\rho\pi}$ describes  the NLO  relativistic and  radiative  corrections, respectively. 
The integral $I^{(1)}$ for $J/\psi$ and  $\psi^{\prime}$ states is the same ( remind that the different leading-order NRQCD matrix elements are taken as the overall normalisation in the Eq.(\ref{AVP})).  Therefore, if $I^{(1)}>0$ and large enough  in order to cancel  the negative  contribution $J^{\text{lo}}+\left\langle {v}^{2}\right\rangle J^{(1)}$  for $\psi^{\prime}$, then this naturally explains the small width for  $\psi^{\prime}$.   It follows from the Table~\ref{Jrp} that the required value of the radiative correction   $\alpha_sI^{(1)}$ must be about $50\%$ of $J_{\rho\pi}^ {\text {lo}}$, which is not that unrealistic given the moderate value of the charm mass.  

The positive contribution $\alpha_s I^{(1)}$ will simultaneously improve the description of  $J/\psi\to \rho\pi$  because it will compensate for the negative effect from the relativistic correction.  Such cancellation  is in agreement with the observation that  the leading-order description  $J/\psi\to \rho\pi$ provides a qualitatively good estimate.  

Taking into account the values of the integrals in Table~\ref{Jrp}  and  other corrections in Eq.(\ref{AVP})  one can conclude that  relativistic corrections strongly  reduce the values of the  branching fractions.  It is clear that resulting values do not describe the data and therefore the possible effect from  the radiative corrections is  very  important for the further progress in the  understanding  of  this  decay.  Therefore, we postpone a detailed phenomenological analysis until we have a complete next-to-leading correction.

\subsection{  $\psi(nS)\to\gamma\pi^0$ decays }
 
The considered analysis can also be applied for other  decays channels $J/\psi\to VP$. The good
feature of the collinear factorisation is that the hadronic nonperturbative
content is described in terms of universal process independent LCDAs. Many of
these functions were already studied in the literature.  Even if the hard kernels are the same the differences in the models for LCDAs can affect the numerical balance and change the value $Q_{hh'}$. 

  Consider, for  example, the decay  of S-wave charmonia into $\gamma\pi^{0}$ final state.  In this case the decay
amplitude is described by the same diagrams as in Fig.\ref{diagrams}(a) but  with the photon LCDAs instead of $\rho$-meson.  These diagrams describe the  photon as a hadron, i.e. such contributions are sensitive to the  nonperturbative components of the photon wave function.    Such  contribution  can provide a sizeable impact, see e.g. discussion in Ref.\cite{Ball:2002ps}. 
 We will refer to this contribution as  hadronic one. 
 
 The contribution with  the perturbative photon coupling  appear  from the diagrams Fig.\ref{diagrams}(b) only
and therefore they are suppressed by electromagnetic coupling $\alpha$, which approximately scales as   $\alpha^3_{s}$.  We will call this contribution electromagnetic.  On the other hand the  hadronic contribution  is suppressed as   $\Lambda^2/m^2_c$ comparing to electromagnetic one.  As a result both contributions can give a comparable numerical effect.  

The data for the branching fractions  $\psi(nS)\to \gamma\pi $ are known \cite{Workman:2022ynf}%
\begin{equation}
\text{Br}[J/\psi\rightarrow\gamma\pi^{0}]=3.56(17)\times10^{-5},\ \ \text{Br}%
[\psi^{\prime}\rightarrow\gamma\pi^{0}]=0.104(22)\times10^{-5},
\end{equation}
which yields%
\begin{equation}
Q_{\gamma\pi}\simeq0.03.
\end{equation}
The width $\Gamma\lbrack J/\psi\rightarrow\gamma\pi^{0}]$ can be well
estimated using data for $\Gamma\lbrack J/\psi\rightarrow\rho\pi^{0}]$ and VDM model
\cite{Chernyak:1983ej}. This indirectly support the picture with the large contribution from  the non-perturbative photon coupling. However, the ratio $Q_{\gamma\pi}$ is  about an order of magnitude larger than $Q_{\rho\pi}$.  Using the results of the previous section we can calculate the hadronic contribution explicitly  and  clarify  the role of the relativistic corrections in this decay. 

The decay amplitude $A_{\gamma\pi}$ is defined similar to $A_{\rho\pi}$ in Eq.(\ref{def:Apv}) with the photon instead of $\rho$-meson.  Now it is given by the sum of two terms
 \begin{equation}
 A_{\gamma\pi}=A_{em}+A_{h},
 \label{Agpi}
 \end{equation}
 which describe electromagnetic and hadronic contributions, respectively.

The leading-order electromagnetic  contribution have been obtained in Ref.\cite{Chernyak:1983ej}.  The relativistic corrections to this amplitude is  similar to one in $J/\psi\to e^+e^-$, see {\it e.g.} Refs. \cite{Bodwin:2002cfe,Bodwin:2007fz}.   The final result  reads
 \begin{equation}
A_{em}=\left\langle 0\right\vert \chi^{\dag}\boldsymbol{\sigma}\cdot\boldsymbol{\epsilon}\psi
\left\vert \psi(n,\boldsymbol{\epsilon})\right\rangle 
\sqrt{2M_{\psi}}  
\frac{ f_{\pi}}{M_{\psi}^{2}}\ e(4\pi\alpha)\ \frac{\sqrt{2}}{9}\ 
(1-f(\left\langle v^{2}\right\rangle ))\int_0^1 du\frac{\phi_{2\pi}(u)}{u},
\label{Aem}
\end{equation}
where
\begin{equation}
f(x)=\frac{1}{3}\frac{x}{(1+x+\sqrt{1+x})}=\frac{x}6+\mathcal{O}(x^2).
\end{equation} 
The leading-order approximation is defined taking $\langle v^{2}\rangle=0$ in Eq.(\ref{Aem}). In this limit $f(\langle v^{2}\rangle)=0$  but we leave unexpanded the quarkonium mass $M_\psi$, which appears in Eq.(\ref{Aem}) from the virtual photon propagator and relativistic normalisation.   

In order to compute the hadronic amplitude $A_h$ we  use  the photon LCDAs from Ref.\cite{Ball:2002ps}. The twist-2
light-cone matrix element is defined as 
\begin{equation}
\left\langle \gamma(q,e)\right\vert \bar{q}(z_{1-})\Dsl{\bar{n}}\gamma_{\bot}^{\mu
}q(z_{2-})\left\vert 0\right\rangle =e_{q}e f_{\gamma}\varepsilon_{\bot}^{\ast\mu
}(qn)\ \int_{0}^{1}dy\ e^{iy(pn)(z_{1}\bar{n})/2+i(1-y)\left(  pn)(z_{2}%
\bar{n}\right)  /2}\ \phi_{2\gamma}^{\bot}(y),
\end{equation}
where $e_{u}=2/3$, $e_{d}=-1/3$, electric charge  $e=\sqrt{4\pi\alpha}$. The model for $\phi_{2\gamma}^{\bot}$  reads \cite{Ball:2002ps}
\begin{equation}
\phi_{2\gamma}^{\bot}(y)\simeq6y(1-y),\ \ \ \ f_{\gamma}(2\text{ GeV}%
)\simeq-47\ \text{MeV}.
\end{equation}
 Twist-3 DAs matrix elements are defined as
\begin{equation}
\left\langle \gamma(p)\right\vert \bar{q}(z_{1-})\Dsl{n}gG_{\mu n}%
(z_{3-})q(z_{2-})\left\vert 0\right\rangle =i e_{q}e f_{3\gamma}%
(qn)^{2}\varepsilon_{\bot}^{\ast\mu}\ \text{FT}\left[  \phi_{3\gamma}%
(y_{i})\right]  ,
\end{equation}%
\begin{equation}
\left\langle \gamma(p)\right\vert q(z_{1-})\Dsl{n}\gamma_{5}g\tilde{G}_{\mu
n}(z_{2-})q(z_{3-})\left\vert 0\right\rangle =-e_{q}e f_{3\gamma}(qn)^{2}\varepsilon_{\bot}^{\ast\mu}\ 
\text{FT}\left[  \tilde{\phi}_{3\gamma
}(y_{i})\right]  ,
\end{equation}
where Fourier transformation is the same as in Eq.(\ref{def:FT}).  The corresponding models for 
  $\phi_{3\gamma}$ and $\tilde{\phi}_{3\gamma}$ was considered in Ref.\cite{Ball:2002ps}
\begin{equation}
\phi_{3\gamma}(y_{i})=360y_{1}y_{2}y_{3}^{2}\left( y_{1}-y_{2}\right)  \ \omega_{3\gamma},\ \
\end{equation}%
\begin{equation}
\tilde{\phi}_{3\gamma}(y_{i})=360y_{1}y_{2}y_{3}^{2}\left(
1+\tilde{\omega}_{3\gamma}\frac{1}{2}\left(  7y_{3}-3\right)  \right)
,\ \ \tilde{\omega}_{3\gamma}\approx\tilde{\omega}_{3\rho}/\zeta_{3}.
\end{equation}
where
\begin{equation}
f_{3\gamma}(2\text{GeV})=- 0.32 \times10^{-2}%
\text{GeV}^{2},\ \ \omega_{3\gamma}\approx\omega_{3\rho},\ \ \ \ \tilde
{\omega}_{3\gamma}\approx\tilde{\omega}_{3\rho}/\zeta_{3}.
\end{equation}
The $\gamma\pi$-decay amplitude can be obtained from Eq.(\ref{AVP})
substituting photon LCDAs instead of $\rho$-meson ones
\begin{equation}
A_{h}  =- \left\langle 0\right\vert \chi^{\dag}\boldsymbol{\sigma}\cdot\boldsymbol{\epsilon}\psi
\left\vert \psi(n,\boldsymbol{\epsilon})\right\rangle
 \sqrt{2M_{\psi}}\left(  \pi\alpha_{s}\right)
^{2}\frac{10}{27}\left(  1+\frac{m}{E}\right)  \frac{f_{\gamma}^{\bot}f_{3\pi
}}{\left[  4E^{2}\right]  ^{2}}\left(  J_{\pi}+\frac{f_{3\gamma}\ f_{\pi}%
}{f_{3\pi}f_{\gamma}}\ J_{\gamma}\right),
\label{Ah}
\end{equation}
where, remind,   $E=m_{c}\sqrt{1+\langle v^{2}\rangle }$.

The ratio of the normalisation couplings in Eq.(\ref{Ah}) yields
($\mu=2\ $GeV)
\begin{equation}
\frac{\ f_{3\gamma}\ f_{\pi}}{f_{3\pi}f_{\gamma}}\approx2.92,\label{Rgamma}%
\end{equation}
which is different from the analogous ratio for the $\rho$-meson (\ref{Rrho}).

To study the effect of the relativistic corrections we again consider three different approximations: leading-order contribution, next-to-leading order contribution (LO + the first correction), and the sum of all powers relativistic corrections as we did for the integrals in Table~\ref{Jrp}.  The results for  the total convolution
hadronic integrals 
\begin{equation}
J_{\gamma\pi}= J_{\pi}+\frac{ f_{3\gamma}\ f_{\pi} } { f_{3\pi} f_{\gamma} }  J_{\gamma},
\end{equation}
  are presented in Table~\ref{Jgp}.
\begin{table}[th]
\caption{Numerical result for the convolution integrals $J_{\gamma\pi}$}
\begin{center}
\label{Jgp}
\begin{tabular} [c]{|c|c|c|c|}
\hline
& $J_{\gamma\pi}^{\text{lo}}$ & $\ J_{\gamma\pi}^{\text{nlo}}/J_{\gamma\pi
}^{\text{lo}}$ & $ J_{\gamma\pi}/J_{\gamma\pi}^{\text{lo}}$\\\hline
$J/\psi$ & $932$ & $0.62$ & $0.45$\\\hline
$\psi^{\prime}$ & $932$ & $-0.49$ & $-0.74$\\\hline
\end{tabular}
\end{center}
\end{table} 
Comparing with the analogous results for the $\rho\pi$-channel one finds   that  the both
descriptions are qualitatively  similar despite  the different ratio
(\ref{Rgamma}) and  the differences between the LCDAs $\phi_{2\gamma}$ and
$\phi_{2\rho}^{\bot}$.  Comparing the different relativistic corrections, one again concludes that the largest numerical effect is provided by the contribution of relative order $v^2$.  However, in the present case, the hadronic contribution gives only a part of the overall result.

 It is also useful to compare the contributions of the different amplitudes in Eq.(\ref{Agpi}) for different charmonium states. 
 In the following numerical estimates  we calculate the NRQCD matrix element  for excited state $\psi'$  in Eq.(\ref{Ah})  using 
 $|R_{20}(0)|^2=0.583\ $GeV$^{3}$ obtained for the Buchmüller-Tye potential in Ref.\cite{Eichten:1995ch}. For the  NLO approximation we perform expansion of the integrals $J_{\pi,\gamma}$ and the factor $(1+m/E)$ in Eq.(\ref{Ah}) but  we  do not expand the factor  $\left[  4E^{2}\right]  ^{2}=[4m_{c}(1+\left\langle v^{2}\right\rangle )]^{2}$  in the denominator. This factor is closely associated with the virtualities of the gluon propagators in the diagram in Fig.\ref{diagrams}(a)  and we assume  that  the quarkonium  mass  $M_{\psi}^{2}\simeq 4m_{c}^{2}(1+\left\langle {v}^{2}\right\rangle )$  is a more natural scale  in this case similar  to the photon virtuality in the amplitude $A_{em}$.  
 \begin{table}[th]
\caption{ Numerical result for the amplitudes $A_{em}$ and $A_{h}$ for various approximations, see more explanations in  the text. The amplitudes have dimension of mass by definition, see Eq.(\ref{def:Apv}), and  their values are presented  in MeV.  }
\label{Amps}
\begin{center}
\begin{tabular}
[c]{|c|c|c|c|c|}\hline
& $A_{em}[J/\psi\rightarrow\gamma\pi^{0}]$ & $A_{h}[J/\psi\rightarrow\gamma
\pi^{0}]$ & $A_{em}[\psi^{\prime}\rightarrow\gamma\pi^{0}]$ & $A_{h}%
[\psi^{\prime}\rightarrow\gamma\pi^{0}]$\\\hline
LO & $0.351$ & $0.514$ & $0.205$ & $+0.425$\\\hline
NLO & $0.338$ & $0.201$ & $0.183$ & $-0.065$\\\hline
Sum & $0.340$ & $0.147$ & $0.190$ & $-0.105$\\\hline
\end{tabular}
\end{center}
\end{table}
These results show that relativistic corrections to $A_{em}$ are relatively small, but  to $A_{h}$ they are large. For excited state they are so large that change the sign of the  hadronic amplitude.  As a result,   the total amplitude for the $\psi\to \gamma\pi$ is much smaller compared to $J/\psi\to \gamma\pi$. 

The presence of the relatively large and  positive amplitude $A_{em}$  makes  less critical the dependence on the  numerical effect from the radiative corrections  therefore  it is interesting to study, at least qualitatively, the  resulting values of the branching fractions. The numerical results for the different approximations are presented in the Table~\ref{Brs}.  In order to get these values we used the total widths $\Gamma_{J/\psi}=93\ $KeV and $\Gamma_{\psi'}=299 $KeV from Ref.\cite{Workman:2022ynf}.
\begin{table}[th]
\caption{ Numerical results for the branching ratios for various approximations. The values of the branchings fractions are given in units $10^{-5}$. }
\label{Brs}
\begin{center}
 \begin{tabular}
[c]{|c|c|c|c|}\hline
& Br$[J/\psi\rightarrow\gamma\pi^{0}]$ & Br$[\psi^{\prime}\rightarrow\gamma
\pi^{0}]$ & $Q_{\gamma\pi}$\\\hline
LO & $10.83$ & $1.50$ & $0.138$\\\hline
NLO & $4.21$ & $0.05$ & $0.012$\\\hline
Sum & $3.43$ & $0.03$ & $0.008$\\\hline
\end{tabular}
\end{center}
\end{table} 

We observe that the LO approximation overestimates the values of the width, but  relativistic corrections reduces these values by 2-3 times for $J/\psi$ and by two orders of magnitude for $\psi'$.  Therefore  resulting values for $J/\psi$ are in relatively good agreement with the data while the values for $\psi'$ are  by factor 2-3 smaller.  But one has to remember  that we have in the background uncalculated  radiative corrections and various uncertainties: unknown higher order relativistic corrections, the choice of normalisation, charm mass, meson LCDAs, {\it etc.}   We postpone the detailed analysis until  the radiative corrections are available.  But let us  notice that  the strong cancellations in the amplitude for the excited state $\psi'$ require a very precise calculation of each term to get a reliable accuracy for the difference.  At the same time the hadronic amplitude has many uncertainties  associated with various sources: relatively large  higher order relativistic corrections, the choice of normalisation, charm mass, meson LCDAs , etc.  Therefore  it seems, that theoretical predictions  for $\psi'$ will have very large errors because of  these uncertainties.

\section{Conclusions}

In conclusion, we calculated  and investigated relativistic corrections  to the decay amplitudes $\psi(nS)\to \rho\pi$  and  $\psi(nS)\to \gamma\pi$
within the pQCD  (NRQCD and collinear factorisation) framework.  This calculation includes the exact correction of relative order $v^2$ and subset of the higher order corrections associated with the quark-antiquarks wave function.   Numerical estimates  show that  an order $v^2$  correction is large and give the dominant numerical effect, which can be  related to the structure of the  collinear integrals.  If this observation  is not affected by other higher order  relativistic corrections, then one has to consider the relative $v^2$-contribution as a special case.  The obtained  relativistic corrections are negative and large.   In case $\psi'\to\rho\pi$  the relative $v^2$ contribution is much larger than the leading-order one.  
Different relativistic correction effects for $J/\psi\to\rho\pi$ and $\psi'\to\rho\pi$ suggest a scenario that may shed light on the $\rho\pi$-puzzle.

 If the QCD radiative correction is positive and large enough, then it will interfere destructively with the relativistic correction for $\psi'\to\rho\pi$, giving a small branching fraction.
 At the same time such radiative correction will improve the description of   $J/\psi\to \rho\pi$ reducing the negative effect of the relativistic correction.  Therefore, we believe that  further investigation of  relative order $v^4$ corrections  and  QCD radiative corrections can  help  to verify  such a scenario.

The same approach  can also be used for an analysis other similar  decay channels.  As a simplest example,   the decay  $\psi(nS)\to \gamma\pi$ is  considered.  In this case a part of the amplitude  is given by similar diagrams but with nonperturbative  photon instead of $\rho$-meson.  Despite the difference between the models for the twist-2 LCDAs,   the qualitative  effect from  the relativistic corrections is quite similar, they are also large and negative.  In this case the  part of the decay amplitude  is described by the electromagnetic subprocess $\psi\to\gamma^*\to \gamma \pi$.  The inclusion of  the relativistic corrections allows to improve the leading-order description.  Again, the large cancellation between the hadronic and electromagnetic contributions  for the $\psi'\to \gamma\pi$  leads to the small branching fraction comparing to   $J/\psi\to \gamma\pi$.  The obtained results show a qualitative agreement with the data.  A calculation of the radiative corrections can also improve the theoretical description in this case too.  
        
\section{Appendix}

Here we provide the analytical expressions for the integrals $J_{\pi}$ and
$J_{\rho}$ introduced  in Eq.(\ref{AVP}). In order to simplify notations  we
use
\begin{equation}
\left\langle v^{2}\right\rangle \equiv\boldsymbol{v}^{2},\ \ \delta
=1-1/\sqrt{1+\boldsymbol{v}^{2}}.\
\end{equation}
The first integral in Eq.(\ref{AVP}) reads
\begin{equation}
J_{\pi}\left(  \boldsymbol{v}^{2}\right)  =\int Du_{i}\ \frac{\phi_{3\pi
}(u_{i})}{u_{1}u_{2}u_{3}}\int_{0}^{1}dy\ \frac{\phi_{2}^{\bot}(y)}{y\bar{y}%
}\left(  \frac{2A_{\pi}}{D_{1}D_{3}}+\frac{B_{\pi}}{D_{1}D_{2}}\right)
,\ \ \bar{y}=1-y,
\end{equation}
where
\begin{equation}
D_{i}=\delta_{i1}\ \left(  y_{1}\bar{u}_{2}+\bar{y}_{1}u_{2}\right)
+\delta_{i2}\ \left(  y_{2}\bar{u}_{1}+\bar{y}_{2}u_{1}\right)  +\delta
_{i3}\ u_{3},
\end{equation}
with
\begin{equation}
y_{1}=y,\ y_{2}=\bar{y}.
\end{equation}
The symbol $\delta_{ik}$  denotes  the  Kronecker delta. 

The numerators  $A_{\pi}$ and $B_{\pi}~$are given by the sums
\begin{equation}
A_{\pi}=\sum_{k=0}^{4}f_{k}^{A}I_{k}[13],\ \ \ B_{\pi}=\sum_{k=0}^{4}f_{k}%
^{B}I_{k}[12],
\label{ABpi}
\end{equation}
where%
\begin{equation}
I_{k}[ij]=\frac{1}{2}\int_{-1}^{1}d\eta\ \frac{\boldsymbol{v}^{k}\eta^{k}%
}{\left(  1+\boldsymbol{v}\eta\ a_{i}\right)  \left(  1-\boldsymbol{v}%
\eta\ a_{j}\right)  }=\frac{\boldsymbol{v}^{k}}{a_{i}+a_{j}}
\sum_{n=0}^{\infty}\boldsymbol{v}^n \frac{a_{j}^{n+1}+(-1)^{n}a_{i}^{n+1}}{n+1+k}\frac{1}{2}\left[
1+(-1)^{n+k}\right]  .\ \ \label{def:Ik}%
\end{equation}
with%
\begin{equation}
a_{j}=\delta_{1j}\left(  1-\delta\right)  \frac{y_{1}-u_{2}}{y_{1}\bar{u}%
_{2}+\bar{y}_{1}u_{2}}+\delta_{2j}\left(  1-\delta\right)  \frac{y_{2}-u_{1}%
}{y_{2}\bar{u}_{1}+\bar{y}_{2}u_{1}}-\delta_{3j}\left(  1-\delta\right)  .
\end{equation}

The coefficients  $f_{k}^{A,B}\equiv f_{k}^{A,B}(u_{i},y;\delta)$ in Eq.(\ref{ABpi})
read%
\begin{equation}
f_{0}^{A}=\frac{ \delta}{2}\left(  3u_{3}-2-\delta\right)  ,\ \ \ \ 
f_{1}^{A}=\frac{\delta}{2}\frac{\left(  1-\delta\right)}{\left(2-\delta\right)}u_{3},
\end{equation}%
\begin{equation}
f_{2}^{A}=\frac{1}{2}\frac{\left(  1-\delta\right)  ^{2}}{\left(2-\delta\right)  ^{2}}
\left(  4-3(2-\delta)u_{3}+ 2\delta(1-\delta)\right)  ,
\end{equation}%
\begin{equation}
f_{3}^{A}=-\frac{1}{2}\frac{\left(  1-\delta\right)  ^{3}}{\left(
2-\delta\right)  ^{2}}u_{3},\ \ f_{4}^{A}=-\frac{1}{2}\frac{\left(
1-\delta\right)  ^{4}}{\left(  2-\delta\right)  ^{2}}.
\end{equation}%
\begin{align}
f_{0}^{B}  & =u_{1}y_{1}+u_{1}y_{2}-\frac{\delta}{2}(u_{1}+u_{2}+y_{1}%
+y_{2}-\delta)\\
& + \frac \delta{2-\delta}
\left(  u_{1}y_{1}+u_{1}y_{2}-(2-\delta)\left(
u_{1}+u_{2}+y_{1}+y_{2}\right)  +(2-\delta)^{2}\right)  ,
\end{align}%
\begin{equation}
 f_{1}^{B}=\frac{1}{2}\frac{\left(  1-\delta\right)  }{\left(  2-\delta\right)}
 \left\{  4(u_{1}y_{1}-u_{2}y_{2})+\delta (u_{1}+y_{1}-u_{2}-y_{2}) \right\}  ,
\end{equation}%
\begin{align}
f_{2}^{B}  & =\frac{1}{2}\frac{\left(  1-\delta\right)  ^{2}}{\left(
2-\delta\right)  ^{2}}\left\{  6\left(  u_{1}+u_{2}+y_{1}+y_{2}\right)
-2\left(  u_{1}y_{1}+u_{1}y_{2}\right)  -8\right.  \\
& \left.  +\delta\left(  4-3\left(  u_{1}+u_{2}+y_{1}+y_{2}\right)  \right)
+2\delta(2-\delta) \right\}  ,
\end{align}%
\begin{equation}
f_{3}^{B}=-\frac{1}{2}\frac{\left(  1-\delta\right)  ^{3}}{\left(
2-\delta\right)  ^{2}}\left(  u_{1}+y_{1}-u_{2}-y_{2}\right)  ,\ \ f_{4}%
^{B}=-\frac{1}{2}\frac{\left(  1-\delta\right)  ^{4}}{\left(  2-\delta\right)
^{2}}.
\end{equation}

The $\rho$-meson integral in Eq.(\ref{AVP}) reads%
\begin{equation}
J_{\rho}=\int_{0}^{1}du\ \frac{\phi_{2\pi}(u)}{u\bar{u}}\int Dy_{i}%
\ \ \frac{1}{y_{1}y_{2}y_{3}}\left(  \frac{2A_{\rho}}{y_{3}D_{2}}%
+\frac{B_{\rho}}{D_{1}D_{2}}\right)  .
\end{equation}
The numerators $A_{\rho}$ and $B_{\rho}$ can be written as 
\begin{equation}
A_{\rho}=\phi_{3\rho}(y_{i})\sum_{k=0}^{4}\left[  f_{k}^{A}\right]
I_{k}[23]+\tilde{\phi}_{3\rho}(y_{i})\sum_{k=0}^{4}\left[  \tilde{f}_{k}%
^{A}\right]  I_{k}[23],
\label{Arho}
\end{equation}%
\begin{equation}
B_{\rho}=\phi_{3\rho}(y_{i})\sum_{k=0}^{4}\left[  f_{k}^{B}\right]
I_{k}[12]+\tilde{\phi}_{3\rho}(y_{i})\sum_{k=0}^{4}\left[  \tilde{f}_{k}%
^{B}\right]  I_{k}[12],
\label{Brho}
\end{equation}
where the integrals $I_{k}$ are defined in Eq.(\ref{def:Ik}) with a bit
different combination $a_{j}$ \
\begin{equation}
a_{j}=\delta_{1j}\left(  1-\delta\right)  \frac{y_{1}-u_{2}}{y_{1}\bar{u}%
_{2}+\bar{y}_{1}u_{2}}+\delta_{2j}\left(  1-\delta\right)  \frac{y_{2}-u_{1}%
}{y_{2}\bar{u}_{1}+\bar{y}_{2}u_{1}}+\delta_{3j}\left(  1-\delta\right)  \ ,
\end{equation}
and we again use for the two-particle LCDA $\ u_{1}=u$, $u_{2}=1-u$.

The coefficients $f_{k}^{A,B}$ and  $\tilde{f}_{k}^{A,B}$ defined in Eqs.(\ref{Arho}) and (\ref{Brho}) read 
\begin{align}
f_{0}^{A}  & =\frac{1}{4}\left(  u_{1}(2y_{3}-\delta)+\delta\left(
\delta-y_{2}-y_{3}\right)  \right)  +
\frac{\delta^2}{ 2}
\\
& \ \ \ \ \ \ \ \ \ \ +\frac{1}{4}\frac{\delta}{(2-\delta)}\left(
u_{1}\left(  6+4y_{3}-3\delta\right)  +\left(  2-\delta\right)  \left(
3y_{2}+2y_{3}-2-3\delta\right)  \right)  ,
\\
f_{1}^{A}  & =-\frac{1}{4}\frac{1-\delta}{2-\delta}\left( 4u_{1}  +\delta  \right) y_{3}  ,
\end{align}%
\begin{equation}
f_{2}^{A}=-\frac{1}{4}\frac{(1-\delta)^{2}}{(2-\delta)^{2}}\left(  
2u_{1}y_{3}+(2-\delta)\left(  2u_{1}+2y_{2}+y_{3} \right)  -2(1-\delta)(2-\delta)\right)  ,
\end{equation}%
\begin{equation}
f_{3}^{A}=\frac{1}{4}\frac{(1-\delta)^{3}}{(2-\delta)^{2}}y_{3},\ \ f_{4}%
^{A}=0.
\end{equation}%
\begin{align}
f_{0}^{B}  & =-f_{2}^{B}=\frac{\delta}{4}\left(  u_{1}-u_{2}-y_{1}+y_{2}\right)  ,\ \\
\ \ f_{1}^{B}  & = -f_{3}^{B}=-\frac{\delta}{4}(1-\delta)\left(  u_{1}+u_{2}-y_{1}-y_{2}\right)
,\ \ \ f_{4}^{B}=0.
\end{align}
\begin{align}
\tilde{f}_{0}^{A}  & =\frac{1}{4}\left(  u_{1}(2y_{3}-\delta)+\delta\left(
\delta-y_{2}-y_{3}\right)  \right)  \\
& +\frac{1}{4}\frac{\delta}{(2-\delta)}\left(
u_{1}\left(  2+4y_{3}-\delta\right)  +\left(  2-\delta\right)  \left(
2+y_{2}-2y_{3}-3\delta\right)  \right)  ,
\end{align}%
\begin{equation}
\tilde{f}_{1}^{A}=-\frac{1}{4}\frac{1-\delta}{2-\delta}\left( 
 4u_{1}y_{3}  +\delta\left(  2u_{1}-2y_{2}+y_{3}\right)  \right)  ,
\end{equation}%
\begin{equation}
\tilde{f}_{2}^{A}=-\frac{1}{4}\frac{(1-\delta)^{2}}{(2-\delta)^{2}}\left(
2u_{1}y_{3}-3y_{3}(2-\delta)+\delta(2-\delta) \right)  ,
\end{equation}%
\begin{equation}
\tilde{f}_{3}^{A}=\frac{1}{4}\frac{(1-\delta)^{3}}{(2-\delta)^{2}}\left(
2u_{1}-2y_{2}+y_{3}\right)  ,\ \ \ \tilde{f}_{4}^{A}=-\frac{1}{2}%
\frac{(1-\delta)^{4}}{(2-\delta)^{2}}.
\end{equation}%
\begin{equation}
\tilde{f}_{0}^{B}=\frac{\delta}{4} \left(  3\left(  u_{1}+u_{2}+y_{1}+y_{2}\right)  -4-2\delta\right), \ \ \ 
\tilde{f}_{1}^{B}=-\frac{\delta}{4}\frac{1-\delta}{2-\delta}\left(  u_{1}-u_{2}+y_{1}-y_{2}\right)  ,
\end{equation}%
\begin{equation}
\tilde{f}_{2}^{B}=\frac{1}{4}\frac{(1-\delta)^{2}}{(2-\delta)^{2}}\left(
8-3\left(  2-\delta\right)  \left(  u_{1}+u_{2}+y_{1}+y_{2}\right)
+4\delta(1-\delta) \right)  ,
\end{equation}%
\begin{equation}
\tilde{f}_{3}^{B}=\frac{1}{4}\frac{(1-\delta)^{3}}{(2-\delta)^{2}}\left(
u_{1}-u_{2}+y_{1}-y_{2}\right)  ,\ \ \ \tilde{f}_{4}^{B}=-\frac{1}{2}%
\frac{(1-\delta)^{4}}{(2-\delta)^{2}}.
\end{equation}

\end{document}